\documentclass[aps,prc,groupedaddress,showpacs,showkeys]{revtex4}
\usepackage{latexsym,epsfig,amsmath,amssymb}
\usepackage{graphicx}
\usepackage{graphics}

\newcommand{\seq}{\begin{subequations}}
\newcommand{\sen}{\end{subequations}}
\newcommand{\eq}{\begin{eqnarray}}
\newcommand{\en}{\end{eqnarray}}

\def\L2{\Lambda^2}

\begin{document}

\title{A discussion of deuteron transverse charge densities}
\author{Cuiying Liang$^{1,2,3}$, Yubing Dong$^{1,2}$, and Weihong Liang$^{3}$
\vspace*{.3\baselineskip}\\}
\affiliation{
$^{1}$ Institute of High Energy Physics, Chinese Academy of Sciences,
Beijing 100049, P. R. China \vspace*{.3\baselineskip}\\
$^{2}$ Theoretical Physics Center for Science Facilities (TPCSF),
CAS, P. R. China \vspace*{.3\baselineskip}\\
and \vspace*{.3\baselineskip}\\
$^{3}$ College of Physical Science and Technology, Guangxi Normal University,
 Guilin 541004, P. R. China\\}
\date{\today}

\begin{abstract}
The deuteron transverse charge density $\rho_C(b)$ is the
two-dimensional Fourier transform of its charge form factor in the
impact space. We show that different parameterizations of the charge
form factors provide different $\rho_C(b)$, in particular at the
central value of impact parameter ($b=0$), although all the
parameterizations can well reproduce the form factors in the region
of small $Q^2$. In addition, we also check the explicit
contributions from the different coordinate intervals of the
deuteron wave function
to its root-mean-square radius.\\
\end{abstract}



\maketitle
%\newpage
\section{Introduction}
\begin{figure}
  \centering
  % Requires \usepackage{graphicx}
  \includegraphics[width=6cm, height=4cm]{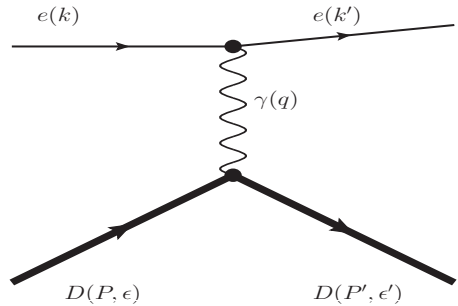}\\
  \caption{Electron-deuteron elastic scattering
  in the one-photon approximation.}
\end{figure}

The deuteron form factors have received much attention and have been
extensively discussed for several decades (for some recent reviews,
see, e.g.~\cite{Gilman,Sick, Gross, Garcon1}). A deuteron, as a spin-1
particle, has three form factors of charge $G_C$, magnetic $G_M$,
and quadrupole $G_Q$. It is often regarded as a loosely bound state
of the proton and neutron (with binding energy $\epsilon_D \sim
2.22$~MeV), and consequently the study of the deuteron properties
can shed light on the structure of the nucleon (in particular of the
neutron) as well as of the nuclear effects. The deuteron
electromagnetic (EM) properties can be explored by the lepton-deuteron
elastic scattering. The matrix element of the electron-deuteron
($eD$) elastic scattering in the one-photon approximation, as
shown in Fig. 1, can be written as
\eq
{\cal M}=\frac{e^2}{Q^2} \bar
u_e(k^\prime) \gamma^{\mu} u_e(k) {\cal J}_{\mu}^D(P,P^\prime),
\en
where $k$ and $k^\prime$ are the four--momenta of initial and final
electrons. ${\cal J}_{\mu}^D(P,P^\prime)$ is the deuteron EM
current, and its general form is \eq\label{D_current} {\cal
J}_{\mu}^D(P,P^\prime) = - \biggl( G_1(Q^2)\epsilon^{\prime
*}\cdot\epsilon-\frac{G_3(Q^2)}{2M_D^2} \epsilon\cdot
q\epsilon^{\prime *}\cdot q \biggr) (P + P^\prime)_{\mu} - G_2(Q^2)
\biggl( \epsilon_{\mu}\epsilon^{\prime *}\cdot q - \epsilon^{\prime
*}_{\mu} \epsilon\cdot q \biggr)~,
\en where $M_D$ is the deuteron mass, $\epsilon$($\epsilon^\prime$) and
$P(P^\prime)$ are the polarization and the four--momentum of the initial (final) deuteron,
and $Q^2=-q^2$ is the momentum transfer square with $q=P^\prime - P$. The three EM form factors
$G_{1,2,3}$ of the deuteron are related to the charge $G_C$,
magnetic $G_M$, and quadrupole $G_Q$ form factors by
\eq
G_C =
G_1+\frac23\tau G_Q\,, \hspace*{.25cm} G_M \ = \ G_2 \,,
\hspace*{.25cm} G_Q = G_1-G_2+(1+\tau)G_3, \hspace*{.25cm}
\en
with
$\tau=\frac{Q^2}{4M_D^2}$.

Since the deuteron is an isoscalar particle, if one only consider
the contribution of the isoscalar vector mesons--$\omega$ and
$\phi$, that the parameterizations of the deuteron form factors can
be written as \cite {Tomasi}
\eq
G_i(Q^2)=N_i
g_i(Q^2)F_i(Q^2),~i=C,Q,M, ~~~~~F_i(Q^2)= 1-\alpha_i-\beta_i+
\alpha_i\displaystyle\frac{m_{\omega}^2}{m_{\omega}^2+Q^2}
+\beta_i\displaystyle\frac{m_{\phi}^2}{m_{\phi}^2+Q^2},
\en
where
$m_{\omega}$ ($m_{\phi}$) is the mass of the $\omega$
($\phi$)-meson. Note that the $Q^2$-dependence of $F_i(Q^2)$ is
parameterized in such form that $F_i(0)=1$, for any values of the
free parameters $\alpha_i$ and $\beta_i$, which are real numbers.
The terms $g_i(Q^2)$ can be written as the functions of the two real
parameters of $\gamma_i$ and $\delta_i$, which are generally
different for each form factor
\begin{equation}
g_i(Q^2)=1/\left [1+\gamma_i{Q^2}\right ]^{\delta_i},
\label{eq:eq12}
\end{equation}
and $N_i$ is the normalization of the $i$-th form factor at $Q^2=0$: \eq
N_C=G_C(0)=1,~~~~~N_Q=G_Q(0)=M^2{\cal Q}_D=25.83,~~~~~
N_M=G_M(0)=\displaystyle\frac{M_D}{M_N}\mu_D=1.714, \en where ${\cal
Q}_D$, and $\mu_D$ are the quadrupole and the magnetic moments of
the deuteron, and $M_N$ is the nucleon mass. It should be reiterated
that this parametrization scheme has been discussed explicitly by
Ref. \cite {Tomasi}.

There are a lot of studies on the EM properties of the nucleon and
deuteron, like their EM form factors, in the literature. Recently,
the nucleon transverse charge density attracts great interest, since
this two-dimensional density can directly relate to the matrix
element of a density operator. It stands for the two-dimensional
Fourier transform of the EM form factor and presents for the density
(in the infinite momentum frame) located at a transverse separation
$b$ (impact parameter) from the center of transverse momentum \cite
{Soper,Ralston, Burkardt,Diehl, Carlson}. The transverse densities
of the pion and nucleon have been discussed extensively. In
particular, the transverse density of pion at central $b$ ($b=0$)
could tell from the different model calculations because theoretical
predictions give different behaviors at $b=0$, even if they all can
well reproduce the EM form factors of the pion \cite{Miller1}.

On the other hand, the root-mean-square $(rms)$ radius of the proton
$R=\langle r ^{2}\rangle^{1/2}$ has also become particular interest
recently. This is because the inconsistence between the measurements
of the $ep$ elastic scattering and of the hyperfine structure of the
muonic hydrogen atom. The proton size is, measured by the $ep$
elastic scattering experiment, $(0.879\pm 0.005\pm 0.004)$ fm
\cite{Bernauer}. However, the $rms$ radius can also modify the
hydrogen energy levels. With the 2P-1S transition energy and the 1S
hyperfine structure of hydrogen, which are well measured with a high
degree of accuracy, the $rms$ radius of the proton can be extracted
and the obtained value is $0.8678\pm 0.0069$ fm \cite {Mohr}.
Similar to hydrogen, one can also get the information of the proton
$rms$ radius from muonic hydrogen. The short lifetime and heavy mass
of the muons make the muonic hydrogen energy levels being more
sensitive to the proton $rms$ radius than the hydrogen. The recent
extracted $rms$ radius of the proton is $0.84184\pm 0.00067$ fm,
which is extracted from the hyperfine structure of the muonic
hydrogen atom \cite{Pohl}. This value is about $5\sigma$ smaller
than the ones from $ep$ elastic scattering and from hydrogen
spectroscopy. After this inconsistence was found, more precise
investigation of the $rms$ radius of the proton becomes necessary.
There are many parameterizations of the proton form factors in the
literature, which all fit the data well. Detailed analysis for the
different parameterizations shows that the contributions of the wave
function in the large distance region is sizable \cite{Sick1}.
Therefore, the $rms$ radius of the proton is also sensitive to the
wave function in the large distance.

In this work, analogous to the pion and the proton targets, the
deuteron transverse charge density $\rho_C(b)$ is analyzed with the
help of several sets of parameterizations of the deuteron form
factors which have been given by Ref. \cite{Tomasi}. It should be
mentioned that those parameterizations are employed to fit the
measured deuteron form factors, and therefore, in the low $Q^2$
region, one expects that those all match the data well. However,
when we make the Fourier transform of the form factors in order to
get the densities in the impact parameter space, the densities in
the region of small $b$ (particularly in the central $b=0$) are
expected to be dominantly affected by the form factors in the large
$Q^2$ region. Therefore, the impact parameter $b$-dependence of the
transverse densities provides more information for the form factors
in the large $Q^2$ region. In addition, the deuteron wave function
of the Paris N-N potential \cite{Lacombe} will be employed to
re-study the $rms$ radius of the deuteron and particularly to see
the relation of the $rms$ radius of the deuteron to the wave
function in the large distance (say $r > 4$ fm) region.

\section{Phenomenological analysis}

The density $\rho_C(b)$ of deuteron in the transverse plane is the
two-dimensional Fourier transform of its EM form factors. The
Fourier transform of the charge form factor is (see the detailed
calculation of the transverse charge density in Refs. \cite
{Miller1,Liang})
\begin{equation}
\rho_C(b)=\frac{1}{(2\pi)^2}\int {d^2Q}G_C(Q^2)
e^{i\vec{q}\cdot \textbf{b}},
\end{equation}
where $Q^2=-q^2=\vec{q}^{~2} > 0$ in the Drell-Yan frame \cite
{Miller1} ($q^+=0$), $b$ is the impact parameter in the
two-dimensional transverse plane. Using Eqs. (4) and (5) with the
parameters given by Ref. \cite{Tomasi}, we get the Tables (I-IV),
where the contributions by the integral from the $Q$ intervals of
$(0-\infty)$, $(0-1)$, $(1-2)$, $(2-10)$, and $(10-\infty)$ GeV are
separately displayed. Fig. 2 gives the estimated two dimensional
transverse charge density $\rho_C(b)$ in the impact $b$-space.  Fig.
2 tells that although all the four parameterizations of the deuteron
charge form factors fit the available data in the small $Q^2$
region well, they provide very different $\rho_C(b)$, particularly
in the small center $b$-value. The scheme I of the parameterizations
gives infinity for $\rho_C(b=0)$ (see Table I) , while other three
give 0.72, 1.42, and 0.90 fm$^{-2}$ in the limit of $b=0$, respectively. The
divergence of the transverse charge densities at $b=0$ is expected
to be related to the parameterizations of
$g_i(Q^2)=1/[1+\gamma_iQ^2]^{\delta_i}$ (see Eq. (5)). If $\delta_i$
is smaller than 1, then, the integral with respect to $Q^2$ (see Eq.
(7)) at $b=0$ limit turns to infinity. Therefore, the $b=0$ behavior
of the transverse charge density is dominantly contributed by the
charge form factor in the large $Q^2$ region, and it determines the
form factor in the large $Q^2$ region. Actually, this feature also
can be seen from Tables (I-IV).

We see that there are remarkable discrepancies between the different
parameterizations at the small value of $b$, particularly at $b=0$,
and the discrepancies become smaller when $b$ increases. When the
$b$ becomes smaller, the contributions from large $Q^2$ region to
the integral of Eq. (7) turn to be more important. This is not
surprising. At large $Q^2$ region which corresponds to small $b$
regime, different model parameterizations of $G_C(Q^2)$ are not
always the same. Some $\rho_C(b)$, at central $b$, have definite
numbers, and some are divergent. Different parameterizations of the
charge form factors provide very different $\rho_C(b)$. We find from
Tables (I-IV) that the contribution to $\rho_C(b)$ of $G_C(Q^2)$ in
the region $10\leq Q\leq\infty$ GeV becomes sizeable when $b$ is
decreasing. In particular, in the region of very small $b$-value,
say less than 0.05 fm, the contribution turns dominant.
Substituting the parameters of different model parameterizations
(the parameters in Tables (I-IV)) into Eqs. (4) and (5), we obtain the
ratios of the estimated $\rho_C(b)$ in the region of $Q = (1 - 2)$
GeV to the one in the whole $Q$ region ($Q = (0 - \infty)$ GeV)£¬
\begin{eqnarray}\displaystyle\nonumber
{\rm Ratio}&=&\frac{\int_0^2\frac{d^2q}{(2\pi)^2}G_C(\vec{q}^{~2})e^{i\vec{q}\cdot\textbf{b}}}
{\int_0^{\infty}\frac{d^2q}{(2\pi)^2}G_C(\vec{q}^{~2})e^{i\vec{q}\cdot\textbf{b}}} \\
&=&\frac{\int_0^2\frac{qdq}{2\pi}G_{C}(\vec{q}^{~2})J_0(qb)}{\int_0^{\infty}\frac{qdq}{2\pi}G_{C}(\vec{q}^{~2})J_0(qb)},
\end{eqnarray}
see Fig. 3. It shows more clearly that when b decreases, the
contribution from large $Q^2$ region increases, while the
contribution from small $Q^2$ region (say $0 < Q < 2$ GeV)
decreases. So a well understanding of the form factor in the large
$Q^2$ region is necessary if one wants to know the transverse charge
density near the central value of $b$ in the impact space. It also
means that the transverse charge density at the central $b$ sheds
light on the form factor in the large $Q^2$ region. Whether the
model parametrization is good or not also can be judged by the
estimated $\rho_C(0)$, particularly in the small $b$ region. So far,
our knowledge of the form factors in the large $Q^2$ regime is
rather limited. More precise experimental data are needed in order
to determine the tail of the form factor in the region.

In the region of moderated $b$ (say $0.5\sim 1$ fm), the estimated
transverse charge density of $\rho_C(b)$ is expected to be
dominated by the form factor in the region of $Q\sim 1$ GeV, where
meson cloud effect on the charge form factors of the proton and
neutron is expected to be important. Conventionally, we consider the
three quark core $| 3q\rangle$ in the proton or neutron, the core is
always located in the central $b$ and the size of the core is
expected to be smaller than $0.5$ fm. When the meson cloud, pion
meson cloud for example, is considered, the proton has the
components of $|(3q)^0\pi^+\rangle$ and $|(3q)^+\pi^0\rangle$ and
the neutron has the components of $|(3q)^+\pi^-\rangle$ and
$|(3q)^0\pi^0\rangle$. Thus, the long positive tail of the proton
transverse charge density comes from the positive pion cloud and on
the contrary, the long negative tail of the neutron transverse
charge density results from the negative charged pion cloud \cite
{Miller2}. According to our estimate, the obtained $\rho_C(b)$ all
have a positive long tail. This is due to the positive proton tail.
\begin{figure}
\centering\includegraphics[scale=0.5]{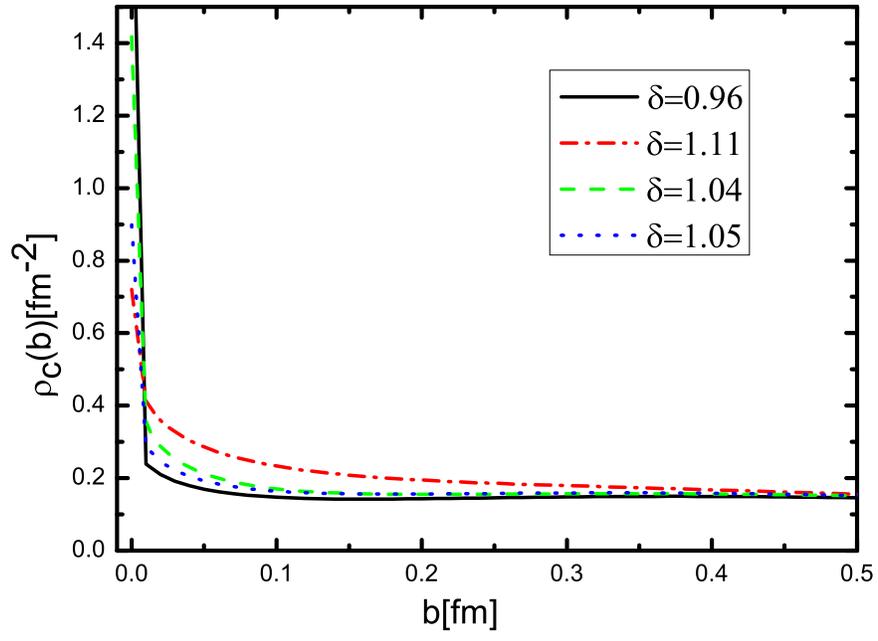} \caption{The
obtained transverse charge densities with four different
parameterizations.}
\end{figure}

\vspace{0.5cm}

\begin{figure}
\centering\includegraphics[scale=0.51]{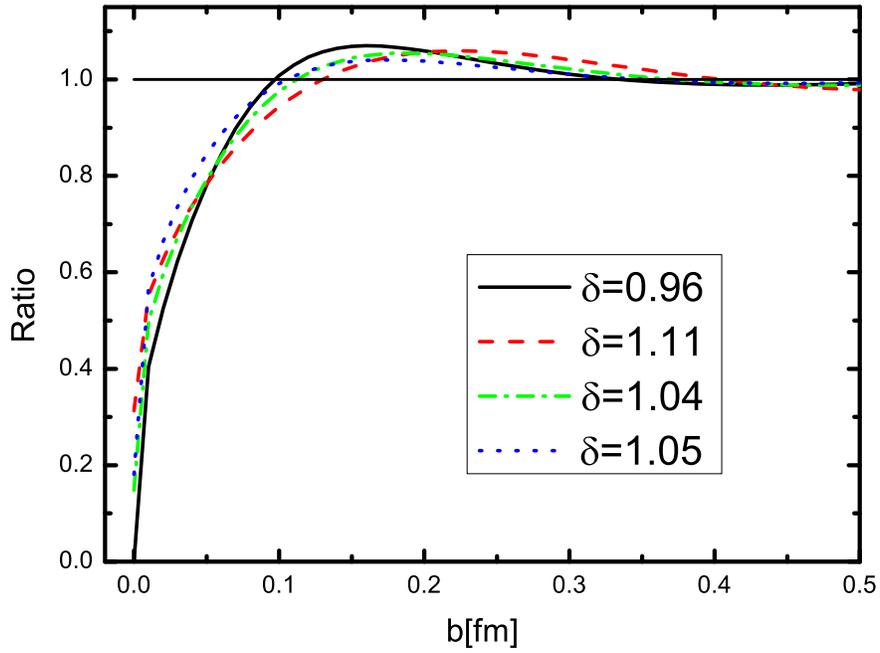} \caption{The ratios
of the transverse charge densities contributed by the region of
$Q=(0-2)$ GeV and by the whole $Q$ region.}
\end{figure}

\begin{table}
\caption{The contributions of different $Q$ ranges to $\rho_C(b)$. The unit of $\rho_C(b)$ is fm$^{-2}$. The
deuteron charge form factor is from the parameterizations of Eqs.
(4--5): $\alpha=5.9$, $\beta=-5.2$, $\gamma=13.9$, $\delta=0.96$ by Ref.
\cite{Tomasi}. The notation $ a\pm n $ stands for $a\times 10^{\pm
n}$.}
\begin{tabular}{r | r |r |r |r |r |r}
\hline\hline
$Q$ (GeV)& $b=0$ fm & $b=0.01$ fm &  $b=0.02$ fm & $b=0.1$ fm & $b=0.5$ fm & $b=1$ fm\\
\hline
$0-\infty$ &$divergent$ & $3.53-01$ & $2.72-01$ &$1.44-01$ & $8.93-02$
& $8.93-02$\\ \hline\hline

$0-1$ &$1.61-01$ & $1.61-01$ & $1.61-01$ & $1.56-01$  & $9.15-02$
& $9.15-02$\\

$1-2$ &$-1.83-02$ & $-1.82-02$ & $-1.82-02$ & $-7.67-03$ & $-2.44-03$
& $-2.44-03$\\

$2-10$& $1.16-01$ & $1.13-01$ & $1.04-01$ & $-6.15-03$ & $1.32-04$
& $1.32-04$\\

$10-\infty$ &$divergent$ & $9.75-02$ & $2.56-02$ & $1.38-03$ & $9.60-05$
& $9.60-05$\\
\hline\hline
\end{tabular}
\label{tab_dwpar}
\end{table}

\begin{table}
\caption{The contributions of different $Q$ ranges to $\rho_C(b)$. The unit of $\rho_C(b)$ is fm$^{-2}$. The
deuteron charge form factor is from the parameterizations of Eqs.
(4--5): $\alpha=5.0$, $\beta=-4.5$, $\gamma=11.5$, $\delta=1.11$ by Ref.
\cite{Tomasi}.The notation $ a\pm n $ stands for $a\times 10^{\pm
n}$.}
\begin{tabular}{r | r |r |r |r |r |r}
\hline\hline
$Q$ (GeV) & $b=0$ fm & $b=0.01$ fm &  $b=0.02$ fm & $b=0.1$ fm & $b=0.5$ fm & $b=1$ fm\\
\hline
$0-\infty$&$7.20-01$ & $4.08-01$ & $3.59-01$ & $2.33-01$ & $1.58-01$
& $8.77-02$\\ \hline\hline
$0-1$  &$2.02-01$ & $2.02-01$ & $2.02-01$ & $2.00-01$ & $1.59-01$
& $8.66-02$\\

$1-2$ & $2.36-02$ & $2.36-02$ & $2.35-02$ & $2.02-02$ & $-7.07-03$
& $1.25-03$\\

$2-10$ &$1.24-01$ & $1.22-01$ & $1.15-01$ & $9.45-03$ & $2.97-03$
& $-2.34-04$\\

$10-\infty$ & $3.71-01$ & $6.14-02$ & $1.88-02$ & $3.76-03$ & $2.98-03$
& $6.91-05$\\
\hline\hline
\end{tabular}
\label{tab1}
\end{table}

\begin{table}
\caption{The contributions of different $Q$ ranges to $\rho_C(b)$. The unit of $\rho_C(b)$ is fm$^{-2}$. The
deuteron charge form factor is from the parameterizations of Eqs.
(4--5): $\alpha=5.75$, $\beta=-5.11$, $\gamma=12.1$, $\delta=1.04$ by
Ref. \cite{Tomasi}. The notation $ a\pm n $ stands for $a\times 10^{\pm
n}$.}
\begin{tabular}{r | r |r |r |r |r |r}
\hline\hline
$Q$ (GeV) & $b=0$ fm & $b=0.01$ fm &  $b=0.02$ fm & $b=0.1$ fm & $b=0.5$ fm & $b=1$ fm\\
\hline
$0-\infty$ &$1.42+00$ & $3.32-01$ & $2.78-01$ & $1.70-01$ & $1.51-01$
& $9.00-02$\\ \hline\hline
$0-1$   &$1.72-01$ & $1.71-01$ & $1.71-01$ & $1.71-01$ & $1.48-01$
& $9.11-02$\\

$1-2$  &$-5.39-03$ & $-5.25-03$ & $-5.23-03$ & $-5.10-03$ & $1.45-03$
& $-1.31-03$\\

$2-10$  &$1.09-01$ & $1.00-01$ & $9.36-02$ & $-3.54-05$ & $1.62-03$
& $1.90-05$\\

$10-\infty$ &$1.14+00$ & $6.64-02$ & $1.89-02$ & $4.28-03$ & $3.25-04$
& $7.60-05$\\
\hline\hline
\end{tabular}
\label{tab1}
\end{table}

\begin{table}
\caption{The contributions of different $Q$ ranges to $\rho_C(b)$. The unit of $\rho_C(b)$ is fm$^{-2}$. The
deuteron charge form factor is from the parameterizations of Eqs.
(4--5): $\alpha=5.50$, $\beta=-4.78$, $\gamma=12.1$, $\delta=1.05$ by
Ref. \cite{Tomasi}. The notation $ a\pm n $ stands for $a\times 10^{\pm
n}$.}
\begin{tabular}{r | r |r |r |r |r |r}
\hline\hline
$Q$ (GeV) & $b=0$ fm & $b=0.01$ fm &  $b=0.02$ fm & $b=0.1$ fm & $b=0.5$ fm & $b=1$ fm \\
\hline
$0-\infty$  &$8.99-01$ & $2.78-01$ & $2.45-01$ & $1.64-01$ & $1.52-01$
& $8.96-02$\\ \hline\hline
$0-1$  &$1.73-01$ & $1.73-01$ & $1.73-01$ & $1.72-01$ & $1.48-01$
& $9.10-02$\\

$1-2$  &$-1.01-02$ & $-1.01-02$ & $-1.00-02$ & $-9.10-3$ & $2.94-03$
& $-1.53-03$\\

$2-10$ &$7.48-02$ & $7.30-02$ & $6.80-02$ & $-1.76-03$ & $9.02-04$
& $7.80-05$\\

$10-\infty$ &$6.61-01$ & $5.15-02$ & $1.46-02$ & $3.07-03$ & $2.33-04$
& $5.40-05$\\
\hline\hline
\end{tabular}
\label{tab1}
\end{table}

In addition, one usually estimates the deuteron root-mean-square
radius from the deuteron model-dependent wave function
\begin{equation}
r_{d}=\frac{1}{2} \Big \{ \int ^{\infty}
_{0}drr^{2}[u^{2}(r)+w^{2}(r)] \Big \}^\frac{1}{2}.
\end{equation}
One gets $r_d=1.966$ fm from the deuteron wave function of Ref.
\cite{Lacombe} . This value results from the parameterized radial
wave function as a discrete superposition of Yukawa-type terms.
%The $r^2_{d}$ can be re-written as:
%\begin{equation}
%r^2_{d}=\frac{1}{4} \int ^{\infty} _{0}drr^{2}\Big [u^{2}(r)+w^{2}(r)\Big ].
%\end{equation}
Here we re-analyze the contributions to the $rms$ radius explicitly.
The obtained $r^2_{d}$ (with D-wave) and $r^2_{ds}$ (without D-wave)
values contributed from the different integral ranges of $r$ are
shown in Table V.
\begin{table*}
\caption{The different integral ranges of $r$ contribute to $r^2_d$ and $r^2_{ds}$.
The notation $ a\pm n $ stands for $a\times 10^{\pm n}$.}
\begin{tabular}{c|c||c|c|c|c|c|c}
\hline\hline
$r$ (fm) &$0-\infty$ & $0-1$   & $1-2$ &  $2-5$ & $5-10$ & $10-20$& $20-\infty$ \\
\hline
$r^2_d$ (fm$^2$)  &$3.855+00$ & $9.890-03 $ & $1.637-01$& $1.341+00$&  $1.720+00$&
$6.107-01$& $1.994-02$ \\
\hline
$r^2_{ds}$ (fm$^2$)  &$3.781+00$ & $9.341-03 $ & $1.509-01$& $1.291+00$&  $1.702+00$&
$6.084-01$& $1.990-02$ \\
\hline\hline
\end{tabular}
\label{tab1}
\end{table*}
We see that the integral, in the region of $r=5\sim 10$ fm, is also
sizeable to $r_d^2$. It means the important role of the long tail of
the wave function. It is clear that the extrapolation deuteron wave
function to the large distance $r$ introduces model dependence and
is less known. Moreover, the S-wave function is active at $2\leq
r\leq 20$ fm and its contribution is dominant then. Finally, in the
$ep$ and $eD$ elastic scattering, different parameterizations or
different model calculations are not the same for the tail of the
wave function in the large distance. Our analysis reiterate that the
observable $r_d^2$ contains the sizeable effect from the integral of
the larger $r$ space. Although different model calculations can all
well reproduce the value of  $r_d^2$, their extrapolated wave
function in the large distance may be different. This phenomenon
also happens in the proton case \cite {Sick1}.

\section{Summary}

To summarize, we simply employed the known phenomenological
parameterizations of the deuteron charge form factors \cite {Tomasi}
to get the transverse charge density of the deuteron in the
two-dimensional impact space. The obtained densities show quite
different behaviors in the limit of $b$=0. This is because the
density in the small $b$ region is related to the form factor in the
large $Q^2$ region and all the parameterizations can well reproduce
the form factor in the small $Q^2$ region, but not clear in the
large $Q^2$ region. Therefore, study of the transverse charge
density can shed light on the form factor in the large $Q^2$ region.
In conclusion, the transverse charge density in the further
measurement is expected to provide the information of the form
factors in the large $Q^2$ region. In addition, we also show that
the model calculation of the $rms$ radius of the deuteron is also
contributed sizeably by the wave function in the large $r$ region.
The extrapolation of the wave function in the large distance,
therefore, is of great interest.

\begin{acknowledgments}

This work is supported by National Sciences Foundations of China
Nos. 10975146, 11035006, 11261130 and 11165005.

\end{acknowledgments}

\end{document}